\documentstyle[preprint,prb,aps]{revtex}
\begin{document}
\author
{
J.P. Badiali
 \\
\vspace{0.5cm}
\textit{LECA, ENSCP-Universit\'e Pierre et Marie Curie,}\\
\textit{4 Place Jussieu, 75230 Paris Cedex 05, France ; 
badiali@ccr.jussieu.fr}\\}

\title { \bf{ENTROPY, TIME-IRREVERSIBILITY AND SCHR\"{O}DINGER EQUATION IN 
A PRIMARILY DISCRETE SPACE-TIME }} \maketitle

\begin{abstract}
In this paper we show that the existence of a primarily 
discrete space-time may be a fruitful assumption from which we may 
develop a new approach of statistical thermodynamics in pre-relativistic 
conditions. The discreetness of space-time structure is determined by a 
condition that mimics the Heisenberg uncertainty relations and the motion 
in this space-time model is chosen as simple as possible. From these two 
assumptions we define a path-entropy that measures the number of closed 
paths associated with a given energy of the
system preparation. This entropy has a dynamical character and depends on 
the time interval on which we count the paths. We show that it exists an 
like-equilibrium condition for which the path-entropy corresponds exactly 
to the usual thermodynamic entropy and, more generally, the usual 
statistical thermodynamics is reobtained. This result derived without 
using the Gibbs ensemble method shows that the standard 
thermodynamics is consistent with a motion that is time-irreversible at a 
microscopic level. From this change of paradigm it becomes easy to derive a 
$H-theorem$. A comparison with the traditional Boltzmann approach is 
presented. We also show how our approach can be implemented in order to 
describe reversible processes. By considering a process defined 
simultaneously by initial and final conditions a well defined stochastic 
process is introduced and we are able to derive a Schr\"{o}dinger equation, 
an example of time reversible equation.

%\\

PACS number:05.90.+m, 05.70.-a, 02.50 -r, 03.65.-w, 47.53+n . 
\end{abstract}
%%%%%%%%%%%%%%%%%%%%%%%%%%%%%%%%%%%%%%%%%%
\vspace{0.5cm}

\section{Introduction}
Despite the large successes obtained in statistical physics during the last 
century we have not been able to derive the second law of thermodynamics 
from statistical mechanics although a large number of works have been 
devoted to this problem (for a review see for instance (\cite{zeh})). In 
the Clausius version, the second law asserts the existence of a state 
function, the entropy, that is a non-decreasing function of time for any 
closed system (\cite{landms}). The difficulty in the derivation of this law 
is the inadequacy between the postulated time-reversible behavior for the 
motion of particles at a microscopic level and the observable irreversible 
behavior of macroscopic systems. In classical mechanics, we have to deal 
with fundamental questions as the existence of the Poincar\'{e}'s 
recurrence time (\cite{zeh}). In quantum physics, Landau and Lifshitz 
(\cite{landmq}) suggested that the origin of irreversibility might be 
related to the measurement process that introduces a difference between 
past and future. Another kind of approach consists to associate the 
thermodynamic arrow of time with the cosmological expansion of the universe 
(see for instance (\cite{schul})). \\ Since the problem of irreversibility 
resists to any demonstration by more than one century it is tempting to 
investigate this problem starting from a new point of view. In a discussion 
about the foundations of statistical thermodynamics in terms of path 
integrals Feynman (\cite{Fain}) developed a given number of fundamental 
remarks leading to ask whether the existence of the hamiltonian is needed 
to formulate classical statistical mechanics $i.e.$ the classical limit of 
quantum statistical physics. In parallel, a possible new foundation of 
statistical mechanics directly in terms of path integral was conjectured. 
The elaborations of a statistical physics without hamiltonian represents a 
new field of investigation (\cite{sork0}). If we adopt this point of view 
we have no more to fight with the Poincar\'{e}'s recurrence time 
(\cite{zeh}) in classical physics but in quantum physics the 
Schr\"{o}dinger equation can not be used as a starting point and the 
canonical form of the density matrix must be abandoned. Consequently new 
physical ingredients have to be introduced but where to find them ?\\ We 
know that the description of the universe at the Planck's scale requires a 
deep modification of our usual physical concepts. For instance, in this 
domain a promising attempt consists to replace point particles by strings 
or to use a discrete space-time structure instead of the differentiable 
manifold of the general relativity ... (\cite{kaku}), (\cite{fink}). It is 
very tempting to see if such new postulates may also change our description 
of the world at a scale much larger than the Planck's one and even in 
pre-relativistic conditions. It is interesting to note that the combination 
of gravitation and quantum physics has led to develop new thermodynamic 
concepts as the holographic principle (\cite{holo}) or the existence of a 
geometric entropy connected with the properties of the quantum vacuum. Such 
an entropy represents a very important discovery that leads to describe 
fundamental thermodynamic laws from vacuum fluctuations without any 
reference to a Gibbs-ensemble description (\cite{schroer}). \\ This paper 
represents an attempt in which we want to show that the existence of a 
primarily discrete space-time may be a useful point of view from which we 
may develop a new approach of statistical thermodynamics. \\ Our approach 
is build up on several steps: \begin{itemize} \item 1. We assume the 
existence of a discrete space-time for which the structure is determined by 
relations that mimic the Heisenberg uncertainty relations. Fortunately, to 
describe usual thermodynamics the continuous limit of this space-time model 
is sufficient. \item 2. We assume that the motion of a particle in 
space-time is as simple as possible. This motion can be characterized by a 
real-valued function describing the transition from one space-time point to 
another. \item 3. Amongst all the quantities that we can introduce to 
characterize the space-time structure and the associated dynamics we choose 
one of them that is similar to the standard thermal entropy. This
path-entropy becomes identical to the thermal one if the energy needed in 
the system preparation equilibrates the mean energy calculated on the 
paths. This result obtained without any reference to the Gibbs ensemble 
method shows that standard thermodynamics is consistent with a motion at a 
microscopic level that is time-irreversible. \item 4. From the previous 
points it is easy to derive a H-theorem without any new assumption in the 
case of free particles. At first glance our derivation of a H-theorem is 
very different from the Boltzmann one nevertheless some comparisons are 
possible.  \item 5. To be really convincing we must also be able to 
describe some situations that we may consider as reversible. Our approach 
is then implemented by the introduction of entry-exit conditions that 
create a past/future symmetry. Then we may describe the system by a complex 
valued function that verifies a Schr\"{o}dinger equation, $i.e.$ a 
time-reversible equation in the Wigner sense. \end{itemize} Note that {\em 
only points 1 and 2 corresponds to real hypothesis.}\\ 
The paper is organized as follows. In Section 2 we introduce 
a space-time model and in Section 3 we describe the motion in this 
space-time. In Section 4 we characterize the space-time by a path-entropy 
and a path-temperature that mimic the correspondent thermodynamic 
quantities. In Section 5 we introduce a like-equilibrium condition from 
which the previous entropy and temperature coincide exactly with the 
corresponding thermodynamic quantities. In Section 6 a H-theorem is 
demonstrated in the case of free particles and a comparison with Boltzmann 
derivation is presented. In the next Section the approach is implemented in 
order to derive a Schr\"{o}dinger equation. In the last Section some 
comments and concluding remarks are presented.

\section{Model of a primarily discrete space-time} 
The choice between a discrete or a continuous version of the space-time 
structure has already been analyzed by Riemann in the classical world and 
more than thirty years ago Feynman presented some doubts concerning the 
continuum nature of space-time in the quantum domain (\cite{fink2}). Today 
it is well accepted that the conventional notions of space and time break 
down at the Planck's scale. Due to this various attempts to elaborate a 
quantum theory based on the existence of a discrete space-time have been 
proposed (see for instance (\cite{fink}), (\cite{sork1}) and (\cite{bruce}) 
for the references quoted therein). \\ 
A discrete space-time means that any length is built up from a finite 
number of the elementary length, $\Delta x$, and any time interval results 
from a series of individual "ticks" of duration $\Delta t$. If a relation 
between $\Delta x$ and $\Delta t$ is expected, its precise form
must depend on the accuracy with which we want to describe the world. To 
establish general aspects of the space-time structure it is possible to 
start from new uncertainty relations issued from string theory (see for 
instance (\cite{Li})) or simple {\em gedanken experiments} (see 
(\cite{naoki}) for a short review in this field). These relations show a 
minimum of uncertainty in the determination of positions that is 
interpreted as the existence of a minimum, $\Delta x$, in the distance 
between two points while $\Delta t$ is the minimum time interval needed to 
characterize two separated points. In the pre-relativistic domain that we 
consider the velocity of light is assumed to be infinite and there is no 
gravitational effect accordingly the only one universal constant that we 
have to consider at the microscopic level is $\hbar$. Thus, if a mass, $m$, 
is located in a region $\Delta x$ the only one relation that we can 
introduce between $\Delta x$ and $\Delta t$ is $(\Delta x)^{2} / \Delta t 
=\hbar /m$. Whatever the values of $\Delta x$ and $\Delta t$ we have 
immediately $\Delta x \Delta p = \hbar$ and $\Delta t \Delta E = \hbar /2$  
provided we use $\Delta p = m\frac{\Delta x}{\Delta t}$ and  $\Delta E = 
(1/2)m (\frac{\Delta x}{\Delta t})^{2}$. {\em Thus our assumption $(\Delta 
x)^{2} / \Delta t =\hbar /m$ is equivalent to suppose that the fine 
structure of space-time is described by relations that mimic the Heisenberg 
uncertainty relations}. This seems a natural choice, it is reasonable to 
expect that the discrete space-time structure contains some ingredients 
that are already familiar to us from our study of the world at a larger 
scale (\cite{sork2}), this leads to a new kind of correspondence principle 
between a discrete space-time and its continuum limit (\cite{raptis}). \\ 
However the relation $(\Delta x)^{2} / \Delta t =\hbar /m$ or the 
uncertainty relations do not fix the precise values of $\Delta x$ and 
$\Delta t$. If we want to avoid particle creation by quantum fluctuations, 
$\Delta x$ must be larger than the Compton wavelength $(\hbar/mc)$ and 
$\Delta t \gg \hbar/ mc^{2}$ where $c$ is the velocity of light; thus, in a 
pre-relativistic world ($c \to \infty$) there is no limitation from below 
for $\Delta x$ and $\Delta t $. Hence we are free to assume that both 
$\Delta x$ and $\Delta t$ tend to zero provided we keep the relation 
$(\Delta x)^{2} / \Delta t =\hbar /m$ in this limit. This is the starting 
point of our approach.

\section{Dynamics in space-time}
Although this is probably not needed, in this first attempt, we 
assume that the space-time points $(t_{i}, x_{i})$ are located on the sites 
of a regular lattice, as in the chessboard problem investigated in 
(\cite{Fain}). Here {\em we assume that the motion is as simple as 
possible}. A particle may jump, at random, from one site to one of its 
nearest neighbors. Thus, by definition, a path corresponds to a set of 
sites $(t_{i}, x_{i})$; the values of ${t_{i}}$ are such as $t_{i+1} > 
t_{i}$ whatever $i$ and the coordinate positions, $x_{i+1}$ is necessarily 
one of the nearest neighbors of $x_{i}$. The conditions over $\Delta x$ and 
$\Delta t$ defined above lead, in the limit $\Delta x, \Delta t \to {0} $, 
to a continuous diffusion process (\cite{itz}) for which the diffusion 
coefficient is $D = \hbar /2m$. This diffusion process has a pure quantum 
origin, $D \to {0}$ if $\hbar \to {0}$. The random walk can be 
characterized by a real-valued continuous function, 
$q_{0}(t_{0},x_{0};t,x)$ representing the density of transition probability 
to go from $(t_{0},x_{0})$ to $(t,x)$ when $t \ge t_{0}$. From 
$q_{0}(t_{0},x_{0};t,x)$ and a function $\phi_{0}( x)$ defined for $t = 
t_{0}$ we form the function $\phi(t,x)$ according to \begin{equation} 
\phi(t,x) = \smallint \phi_{0}(y) q_{0}(t_{0},y;t,x) dy \label{fi1} 
\end{equation} which is the solution of the diffusion equation 
\begin{equation} -{\partial  \phi(t,x) }/{ \partial t} + D \Delta_{x} 
{\phi(t,x)}= 0 \label{diffree} \end{equation} verifying the initial-value 
problem $\phi(t_{0},x) = \phi_{0}( x)$. Note that $q_{0}(t_{0},x_{0};t,x)$ 
is the fundamental solution of (\ref{diffree}) in which $\Delta _{x}$ is 
the laplacian operator taken at the point $x$. In presence of an external 
potential, $u(t,x)$, we generalize (\ref{diffree}) into \begin{equation} 
-{\partial  \phi(t,x)}/{ \partial t} + {D \Delta \phi(t,x)} 
-\frac{1}{\hbar}{u(t, x) \phi(t,x)} = 0 
\label{dif} 
\end{equation} 
In contrast with (\ref{diffree}), the fundamental solution of (\ref{dif}),
$q(t_{0},x_{0};t,x)$, cannot be normalized in general (\cite{naga1}). Thus, 
$q(t_{0},x_{0};t,x)$ is no more a transition probability density but it 
verifies the Chapman-Kolmogorov law of composition (\cite{naga1}) and 
therefore it can be used to describe transitions in space-time. By using 
the Feynman-Kac formula, the fundamental solution of (\ref{dif}) can be 
written in terms of path integral. Then $q(t_{0},x_{0};t,x)$ appears as a 
weighted sum of all the paths $x(t)$ connecting the space-time points 
$(x_{0},t_{0})$ to $(x,t)$ ; the weight of a path is determined by 
 \begin{equation} A[x(t); t, t_{0}] =  \int\limits_{t_{0}}^{t}  
 [\frac{1}{2} 
m [\frac{dx(t')}{dt'}]^{2} + u(t',x(t'))] dt'. 
\label{ACT}  
\end{equation}
and we have 
\begin{equation} q(t_{0},x_{0};t,x) = \smallint 
Dx(t) \exp -\frac{1}{\hbar} A[x(t); t, t_{0}]
\label{q0} 
\end{equation}
Note that the integrand in (\ref{ACT}) looks like the hamiltonian for a free 
particle in presence of an external potential and therefore 
$A[x(t);t,t_{0}]$ will be called the hamiltonian action. However, $A[x(t); 
t, t_{0}]$ is a formal writing (\cite{wiegel}), to calculate the path 
integral we have to discretize $A[x(t); t, t_{0}]$ and the paths that 
contribute to the integral are those for which there is no derivative 
$i.e.$ no velocity in the usual sense (\cite{jpb}). \\

\section{Path-entropy}
{\em A priori}, to describe structure and dynamics in our space-time model 
we may introduce a lot of functions. However, since our main goal is to 
recover thermodynamics we focus on quantities that resemble as far as 
possible the thermodynamic ones. To describe 
equilibrium situations we restrict the external potential, $u(t, x)$, to be 
a time-independent quantity, $u(x)$. Accordingly $A[x(t);t,t_{0}]$ becomes 
$A[x(t); \tau]$ with $\tau = t - t_{0}$. By analogy with the entropy that 
is the key quantity in thermodynamics we have to introduce a quantity that 
measures the order in space-time. Since the standard entropy is defined for 
given values of internal energy and volume we must consider that our 
space-time system is prepared with a given energy $U$ and occupies a volume 
$V$.\\ To characterize the space-time order, around each point $x_{0}$ we 
count the number of paths for which $x(\tau) = x_{0}$ and on which the 
hamiltonian action, $A[x(t); \tau]$, does not deviate too much from the 
action $\tau U$. This is realized by introducing the measure $ exp{- 
\frac{1}{\hbar} [A[x(t); \tau] - \tau U] }$. By analogy with the standard 
thermodynamics we define a path-entropy, $S_{path}$, according to  
\begin{equation} S_{path} = k_{B} \ln \smallint dx_{0} \smallint Dx(t) 
\exp{ - \frac{1}{\hbar} [A[x(t); \tau] - \tau U] }. \label{gam} 
\end{equation} 
Larger is $S_{path}$ larger is the number of paths, this means that 
larger are the acceptable fluctuations of the hamiltonian action around 
$\tau U$ and less strict is the space-time order. Thus $S_{path}$ can be 
used to characterize the order in space-time. However $S_{path}$ depends 
on $\tau$ that is, for the moment, a free parameter and due to this we 
can not conclude that $S_{path}$ corresponds to the thermodynamic entropy. 
$S_{path}$ can be also rewritten as \begin{equation} S_{path} = \frac{k_{B} 
\tau}{\hbar} U + k_{B} \ln Z_{path} \label{entpath} \end{equation}  with 
\begin{equation} Z_{path} = \smallint dx_{0} \smallint Dx(t) \exp{ - 
\frac{1}{\hbar} A[x(t); \tau]} = \smallint dx_{0}q(0,x_{0}; \tau, x_{0}) 
\label{zpath} \end{equation} in which we have taken $t_{0} = 0$ and 
$q(0,x_{0}; \tau, x_{0})$ is the fundamental solution of (\ref{dif}) for 
closed paths and a time interval $\tau$. $Z_{path}$ is the total number of 
closed paths that we may count during $\tau$ irrespective the value of $U$. 
Now we may choose $\tau$ as a function of $U$. \\ To $S_{path}$ we may 
associate a path-temperature $T_{path}$ by $\frac {dS_{path}}{dU} = 
\frac{1}{T_{path}}$ that relies a change of $U$ to a change of $S_{path}$. 
Similarly to $S_{path}$, $T_{path}$ is a well defined quantity but it can 
not be identified with the thermodynamic temperature, at this level. From 
(\ref{entpath}) we have immediately \begin{equation} \frac {1}{T_{path}} = 
\frac{k_{B} \tau}{\hbar} +\frac{k_{B}}{\hbar}[U + \hbar \frac{d}{d \tau} 
\ln Z_{path}] \frac{d \tau}{dU} \label{defto} \end{equation} To calculate 
$\frac{d}{d\tau} \ln Z_{path}$ we use the expression of $Z_{path}$ in term 
of $q(0,x_{0}; \tau,x_{0})$ given in (\ref{zpath}) and  from the 
Chapman-Kolmogorov law of composition we may write \begin{equation} 
q(0,x_{0}; \tau,x_{0}) = \smallint dx_{b} q(0,x_{0};\delta t,x_{b}) 
q(\delta t,x_{b};\tau -\delta t, x_{0}) \label{compo1} \end{equation} 
for any $\delta t$ such as $0 < \delta t < \tau$. From 
(\ref{zpath}), (\ref{compo1}) and (\ref{dif}) we have 
\begin{equation} 
\hbar \frac{d}{d \tau} \ln Z_{path} = \frac{\hbar}{Z_{path}} \smallint 
dx_{0} \smallint dx_{b} q(0,x_{0};\delta t,x_{b}) [D \Delta_{x_{0}} 
-\frac{1}{\hbar}u(x_{0})] q(\delta t,x_{b};\tau -\delta t, x_{0}) 
\label{ddto} 
\end{equation} 
The term associated with the external potential can be written 
\begin{equation} 
\frac{1}{Z_{path}} \smallint dx_{0} u(x_{0}) q(0,x_{0};\tau, x_{0}) =  
\smallint dx_{0} <u_{P}(x_{0})>_{path} 
\label{potpath} 
\end{equation} 
We may interpret $<u_{P}(x_{0})>_{path}$ as the potential energy at 
the point $x_{0}$ calculated as an average over the paths. To calculate the 
contribution of the laplacian in (\ref{ddto}) we use the fact that 
(\ref{compo1}) holds if $\delta t \to 0$, in that case $\Delta_{x_{0}}$ 
operates only on the the term which looks like the kinetic energy in 
$A[x(t); \tau]$. A simple calculation gives
\begin{equation}  
\frac{1}{Z_{path}} \smallint dx_{0} \smallint d(\delta x)  
q(0,x_{0};\delta t, x_{0}+\delta x) [ \frac{m}{2}(\frac{\delta x }{\delta 
t})^{2} - \frac{\hbar}{2 \delta t}] q(\delta t,x_{0} +\delta x;\tau -\delta 
t, x_{0}) 
\label{ecpath} 
\end{equation} 
that we can rewrite as 
\begin{equation} \smallint dx_{0}[\frac{m}{2} <(\frac{\delta x }{\delta 
t})^{2}>_{path} - \frac{\hbar}{2 \delta t}] 
\label {kin} 
\end{equation} 
This expression represents the difference between the kinetic energy 
calculated as an average over the paths and the quantum fluctuation of 
energy corresponding to a time interval $\delta t$ on which the kinetic 
energy is calculated. In the limits $\delta x \to 0$ and $\delta t  \to 0$ 
we may replace $q(0,x_{0};\delta t, x_{0}+\delta x)$ in (\ref{ecpath}) by 
the free particle approximation that only depends on $\delta t$ and $\delta 
x$. Moreover we may approximate $q(\delta t,x_{0} +\delta x;\tau -\delta t, 
x_{0})$ by $q(0,x_{0};\tau, x_{0})$ and its integral over $x_{0}$ gives 
$Z_{path}$ as shown in (\ref{zpath}). In the limits that we consider it is 
easy to see that $\frac{m}{2} <(\frac{\delta x }{\delta t})^{2}>_{path} = 
\frac{\hbar}{2 \delta t}$. We may also check this result by performing an 
exact calculation but using an explicit form of $u(x)$. This 
result is also expected from the properties of the initial lattice on which 
we have $\frac{1}{2}m (\frac{\delta x}{\delta t})^{2} = 
\frac{\hbar}{2 \delta t}$ Thus the term in bracket in (\ref{kin}) is well 
defined in the limit $\delta t \to 0$. More generally, we write 
\begin{equation} 
\frac{m}{2} <(\frac{\delta x }{\delta t})^{2}>_{path} = \frac{\hbar}{2 
\delta t} - <u_{K}(x_{0})>_{path} 
\label {kin2} 
\end{equation} 
in which $<u_{K}(x_{0})>_{path}$ is a well behaved function in the limit 
$\delta t \to 0$. In principle $<u_{K}(x_{0})>_{path}$ and 
$<u_{P}(x_{0})>_{path}$ can be calculated for a given potential 
$u(x)$ and a given value of $\tau$. Finally, we can rewrite (\ref{defto}) 
according to \begin{equation} 
\frac{\hbar}{k_{B} T_{path}} = \tau + [ U - \smallint dx_{0} 
(<u_{K}(x_{0})>_{path}+ <u_{P}(x_{0})>_{path})]\frac{d \tau}{dU} 
\label{to} 
\end{equation} 
This equation establishes a relation between $\tau$ and $T_{path}$ 
when $u(x)$ and $U$ are given .\\ 

\section{Thermodynamics}
The square bracket in (\ref{to}) contains $U$, the specific properties of 
the system via the averages over the paths and the variable $\tau$. We are 
free to choose a particular value, $\tau^{*}$, of $\tau$ in such a way that 
the quantity \begin{equation} 
\smallint dx_{a} [<u_{K}(x_{a})>_{path}+ <u_{P}(x_{a})>_{path}] 
\label{upath} 
\end{equation} 
that looks like the internal energy calculated over the paths is indeed 
equal to the energy of the system preparation i.e. $U$, {\em this is 
clearly an equilibrium condition since the energy that is brought to the 
system, $U$, is transformed into the sum of the mean kinetic and potential 
energy associated with the paths}. Thus, whatever the value of the 
derivative $\frac{d \tau^{*}}{dU}$ from (\ref{to}) we get $\tau^{*} 
=\frac{\hbar}{k_{B} T^{*}} = \beta^{*} \hbar$ that fixes the value of 
$T^{*}$ when $\tau^{*}$ is given. The partition function (\ref{zpath}) can 
be rewritten as 
 \begin{equation} 
Z_{path}^{*}= \smallint dx_{a} \smallint Dx(t) \exp{ - \frac{1}{\hbar}} 
\int\limits_{0}^{\beta^{*} \hbar}  [\frac{1}{2} m [\frac{dx(t')}{dt'}]^{2} + 
u(x(t'))] dt'.  
\label{zpath2} 
\end{equation}
This expression is identical to the one of the  standard partition function 
$Z$ obtained in (\cite{Fain}) starting from the canonical form of the 
density matrix provided $T^{*}$ is identified with the thermal temperature 
$T$. This is justified because our definition of $S_{path}$ and $T^{*}$ are 
in agreement with the zeroth law of thermodynamics. In order to prove this 
point we follow the Callen derivation of the zeroth law (\cite{callen}). 
Let consider a system formed by two independent subsystems $S_{1}$ and 
$S_{2}$, they may differ both by the mass and the external potential and 
their preparation energy corresponds to $U_{1}$ and $U_{2}$. The 
equilibrium condition defined above can be written for the global system 
and for each subsystem taken separately. With our definition of $S_{path}$ 
it is easy to see that the entropy is the sum of entropies of each 
subsystem. Now for a fixed value of $U = U_{1} + U_{2}$ we may imagine a 
virtual energy transfer $\delta U$ from one system to the other one. The 
requirement that, at the equilibrium, the total entropy is maximum relative 
to any $\delta U$ leads to the conclusion that the two subsystems must have 
the same value for $T^{*}$, in agreement with the zeroth law. 
The equilibrium condition also leads to the relation 
$\frac{d}{d \beta}^{*} \ln Z_{path}^{*}= - U$ which is expected from 
standard statistical mechanics (\cite{landms}). If we define $F$ according 
to $F = -k_{B}T \ln Z$ we see that (\ref{entpath}) is nothing else than the 
standard relation $F = U - TS_{path}^{*}$ and consequently $S_{path}^{*}$ 
corresponds to the traditional thermal entropy. \\ Thus we have obtained an 
equivalent description of the thermodynamics directly in terms of paths 
without using Gibbs ensemble method. In this dynamic approach the motion 
associated with paths is not due to a hamiltonian but results from the 
primarily discreetness of the space-time. However our approach is more than 
a simple alternative description of thermodynamics as we shall see below. 
(To be short in what follows we drop the uperscript *) \\ 
In standard path integral formalism $\beta \hbar$ is considered as a {\em 
formal time} (\cite{Fain}). Here, $\tau$ results from a combination of 
dynamics determined by (\ref{dif}) and thermodynamics via a particular 
choice in the solution of (\ref{to}), therefore $\tau$ must have a strong 
physical meaning. From standard textbooks in statistical mechanics 
(\cite{landms}) it is well known that there is no entropy on a short period 
of time. To have thermodynamics we must consider time intervals such that 
the quantum fluctuations do not exceed the order of magnitude of the 
typical thermal energy. In the case of free particles the mean value of the 
kinetic energy is $1/ 2 \beta$ and from the time-energy uncertainty 
relation $(\tau / 2 \beta = \hbar /2)$ we have $\tau = \beta \hbar$;   we 
see that $\tau$ represents the relaxation time that we have to wait in 
order to relax the quantum fluctuations and to reach the thermal regime. 
Our derivation of $\tau$ is more general since independent of the external 
potential but it has the same physical meaning. In 
parallel, using the results given in (\cite{jpb}), it is easy to see that 
the paths are located, in average, on a given volume in space; in absence 
of external potential this volume is a sphere of radius $\Lambda = 
\frac{\hbar}{(mk_{B}T)^{1/2}}$ corresponding to the thermal de Broglie 
wavelength, an expected result. However the most important result of our 
approach is that {\em equilibrium thermodynamics is consistent with a 
motion that is time-irreversible at the microscopic level since given by 
(\ref{dif})}. The interest of this result is to open a door from which we 
may establish a $H-theorem$ without any new assumption.

\section{Derivation of a $H-theorem$}
In the Boltzmann approach of the $H-theorem$ (\cite{zeh}) a system of 
particles is first prepared in a non-equilibrium state by external 
constraints. At the time $t =t_{0}$ these constraints are removed and the 
system relaxes towards its equilibrium state. The Clausius version of the 
second law of thermodynamics asserts that during the system relaxation 
it exists a function that increases monotonically versus 
$(t-t_{0})$ and tends to the thermal entropy when $(t-t_{0})$ becomes 
infinite. In the Boltzmann work there is no external potential for 
$t \ge t_{0}$, the relaxation is driven by collisions between particles. 
Hereafter we mimic a similar description in our space-time model.\\ 
Using an external potential we prepare the system in such a way 
that, at the initial time, $t =t_{0}$, the space points are weighted by a 
distribution function, $\phi_{0}(x)$, which is positive and normalized. Any 
quantity calculated in these conditions depends on $\phi_{0}(x)$, at least 
for $(t - t_{0}) \ge 0 $ but finite. For $t \ge t_{0}$ the external 
potential is switched off, $u(x)=0$, and the time evolution of $\phi(t,x)$ 
is given by (\ref{diffree}). In this case $q(t_{0}, x_{0}; t,x)$ is a 
density of transition probability for $t > t_{0}$ and $\phi(t,x)$ defined 
according to (\ref{fi1}) is the probability to be in $x$ at time $t$.\\ 
Now, we introduce a function $H(t)$ that generalizes $S_{path}$ given by 
(\ref{gam}). After relaxation, the space is uniform since $u(x) =0$ 
and the entropy must be given by (\ref{gam}) that we can rewrite 
as $S = k_{B} \ln (V \gamma(\tau))$ where $V$ is the volume of the system. 
During the relaxation process, at each time, $t$, and for each point, $x$, 
we count the number of closed paths that we can form between the instants 
$t$ and $t+ \tau$, this number is still $\gamma(\tau)$. However, the total 
number of paths at this time and this position is $ \gamma(\tau) 
/\phi(t,x)$, it reduces to  $\gamma (\tau) V $ if $\phi(t,x) = 1/V$ $i.e.$ 
in the case of a uniform system. We define the average, 
$S_{total}(t)$, of $ln [\gamma(\tau) /\phi(t,x)]$ taken over the overall 
volume as  
\begin{equation} S_{total}(t) = k_{B} \smallint \phi(t,x) ln 
[\frac{\gamma (\tau)}{\phi(t,x)}] dx 
\end{equation} 
Now we consider the quantity $H(t)= (1/ k_{B}) (S_{total}(t)-S)$ defined by 
\begin{equation}  H(t)= - \smallint \phi(t,x) ln [\phi(t,x)] dx - 
ln[V] \label{H} 
\end{equation} 
where we have used the normalization of $\phi(t,x)$. The first term in 
(\ref{H}) is positive since the integrand is negative due to the fact that  
$(0 \le \phi(t,x) \le 1)$. Thus, $H(t)$ represents the competition between 
two quantities of opposite sign. The solution of (\ref{dif}) in a volume $V$ 
has been given in (\cite{loch}); in the limit $(t - t_{0}) \to \infty$ it 
has been shown that $\phi (t,x) = 1/V $ and consequently $H(t) \to 0$. In 
addition, we can show that (\cite{loch}) \begin{equation} \frac 
{dH(t)}{dt}= (\frac{\hbar}{2m}) \smallint \frac{1}{\phi(t,x)}(\frac{d 
\phi(t,x)}{dx})^{2}dx \quad \ge {0} \label{ThH} 
\end{equation} 
Thus, $H(t)$ is a {\em monotonic increasing} function of $t$ and $H(t)$ 
vanishes in the stationary regime obtained in the limit $(t - t_{0})$ goes 
to infinite. When the time is running, $S_{total}(t)$ increases in a 
monotonic manner to reach a stationary value corresponding to the thermal 
entropy $S$. Thus {\em a $H-theorem$ is demonstrated.} \\ 
The irreversible behavior results from the existence of a space-time, 
which is discrete at a fundamental level and induces a special dynamics not 
connected with a hamiltonian. Nowhere we consider collisions between 
particles as in the Boltzmann dynamics but we investigate the paths 
performed by one mass. To characterize the dynamics on these paths we select 
a given point $(x_{ i}, t_{ i})$ and for a time interval $\delta t \le 
\tau/2$ we consider the forward and backward velocities defined respectively 
as: $V_{ +} = [(x_{i + 1}- x_{ i} ) / \delta t]$ and $V_{-} = [(x _{i}- x_{ 
i - 1}) / \delta t]$ To the product $V_{+} V_{-}$ we associate (\cite{jpb}) 
an average over paths $< V_{+} V_{-} >$; it has been shown that (\cite{jpb}) 
\begin{equation} < V_{+} V_{-} > = <\frac{(x_{i + 1}- x_{ i} )}{\delta t} 
\frac {(x _{i}- x_{ i - 1})}{\delta t} > = - 1 / ( m \beta ) 
\label{corv} 
\end{equation} 
This result, independent of $\delta t$, leads to several remarks. {\em 
First}, for trajectories on which there is a velocity in the usual sense $< 
V_{+} V_{-}> $ is a positive quantity, here the negative value of $< V_{+} 
V_{-} > $ is the signature of the fractal character of the paths. 
{\em Second}, to the negative value of $< V_{+} V_{-} > $ we may associate a 
{\em formal collision}, which reverses the direction of the velocity, in 
average. From (\ref{corv}) we see that the correlations are not destroyed 
after a formal collision in contrast with the molecular chaos hypothesis 
used by Boltzmann (\cite{zeh}). It is only in the limit $T=0$ that the 
velocities $V_{ +}$ and $V_{-}$ become statistically independent. 
{\em Third}, a formal collision appears at any time and everywhere in 
space; in opposite to the Boltzmann dynamics we cannot consider two 
separated time scale, one for the duration of collisions and another one for 
time between collisions.\\
Of course to be very satisfying our approach must also describes some 
situations that we may consider as reversible.

\section{Time-reversible processes and the Schr\"{o}dinger equation}
In the previous Sections we have considered the real-valued function 
\begin{equation} 
\phi(t,x) = \smallint \phi_{0}(y) q(t_{0},y;t,x) dy 
\label{fi2} 
\end{equation} 
that is the solution of the diffusion equation (\ref{dif}) verifying the 
initial-value problem $\phi(t_{0},x) = \phi_{0}( x)$; in (\ref{fi2}), 
$q(t_{0},x_{0};t,x)$ is the fundamental solution of (\ref{dif}). This 
description is quite natural; we know the initial condition for $t=t_{0}$ 
and describe the system for $t \ge t_{0}$. However, it is possible to 
consider another function $\hat \phi(t,x)$ defined according to 
\begin{equation} 
\hat \phi(t,x) = \smallint \ q(t,x;t_{1},y) \hat 
\phi_{1}(y)dy \label{hatfi} 
\end{equation}
for $t_{0} \le t \le t_{1}$. Using the properties of the fundamental 
solution it is easy to prove that $\hat \phi(t,x)$ is the solution of the 
differential equation 
\begin{equation} 
\frac{\partial  \hat \phi(t,x)}{ 
\partial t} + {D \Delta \hat \phi(t,x)} -\frac{1}{\hbar}{u(t, x) \hat 
\phi(t,x)} = 0 
\label{hatdif} 
\end{equation}
verifying the final condition $\phi(t_{1},x) = \hat \phi_{1}( x)$. Up to 
now the use of (\ref{dif}) {\em or} (\ref{hatdif}) is a question of 
convenience depending on the information we have on the system.\\ 
In what follows we consider a new physical situation in which the system is 
defined simultaneously by (\ref{dif}) {\em and} (\ref{hatdif}) $i.e.$ by a 
dynamics defined at an initial and a final time, $t_{0}$ and $t_{1}$ 
respectively. Of course, for {\em the entry-exit conditions} $(\phi_{0}(x), 
\hat \phi_{1},(x))$ our problem is now to predict what happens for 
any $t$ between $t_{0}$ and $t_{1}$. This is reminiscent of the langragian 
mechanics that is know to describe reversible processes. However in our 
case we must show that the entry-exit conditions introduce a well 
identified mathematical object as, for instance, a well defined stochastic 
process. The proof of that has been given by Nagasawa (\cite{naga1}) and we 
will not reproduce the mathematical details here but we will focus mainly 
on the physical aspects.\\ 
The functions $\phi(t,x)$ and $\hat \phi(t,x)$ that describe the transition 
in space-time are not transition probability density and, up to 
now, there is no restriction on $\phi_{0}(x)$ and $\hat 
\phi_{1}(x)$. Hereafter we assume that $\phi_{0}(x)$ and $\hat \phi_{1}(x)$ 
are two non-negative real-valued functions that we choose such as 
\begin{equation} \smallint \phi_{0}(y) dy q(t_{0},y ;t_{1},x) dx  \hat 
\phi_{1}(x) = 1 \label{norm2} \end{equation} From the set $[\phi_{0}(x), 
\hat \phi _{1}(x), q(s,y;t,x)]$, the condition (\ref{norm2}) and
the Chapman-Kolmogorov law of composition, we can show that 
(\cite{naga1})  \begin{equation} \mu(t,x) = \phi(x) \hat \phi(x) 
\label{defmu} \end{equation}
is a non-negative and normalized quantity; we may consider $\mu(t, x)$ as 
the probability distribution of a stochastic process. Moreover, from the 
same ingredients as above we can define a probability measure noted $Q = 
[\phi_{0}q>><<q \hat \phi_{1}] $ (\cite{naga1}). It is possible to put $Q$ 
in a more traditional form by introducing the new variables 
\begin{equation} 
p(s,y;t,x) = \phi (s,y) q(s,y;t,x) \frac{1}{\phi (t,x)} \quad if \quad \phi 
(t,x) \neq {0} \quad and \quad p(s,y;t,x) =0 \quad otherwise. \label{hat} 
\end{equation} and \begin{equation} 
\hat p(s,y;t,x) = \frac{1}{\hat \phi(s,y)} q(s,y;t,x) \hat 
\phi(t,x) \quad if \quad \hat \phi(s,y) \ne 0 \quad and \quad \hat 
(p(s,y;t,x) =0 \quad otherwise \label{forw} 
\end{equation} 
From their definitions, we can see that $p(s,y;t,x)$ and $\hat p(s,y;t,x)$ 
are positive and normalized according to 
\begin{equation} 
\smallint p(s,y;t,x)dy =1 \quad and \quad \smallint  dx \hat p(s,y;t,x) = 
1 \label{norm} 
\end{equation} 
Consequently we can consider $p(s,y;t,x)$ and $\hat p(s,y;t,x)$ as 
transition probability densities. The probability measure $Q$ can be 
rewritten as $Q= [\phi_{0} \hat \phi_{0} \hat p>> $ that is the traditional 
Kolmogorov representation of a Markov process with the initial distribution 
$\mu (t_{0},x) =\phi_{0}(x) \hat \phi_{0}(x)$ and the transition 
probability density $\hat p(s,y;t,x)$. $Q$ is also identical to a time 
reversed Markov process $Q = << p \phi_{1} \hat \phi_{1}]$. Thus the 
probability measure noted $Q = [\phi_{0}q>><<q \hat \phi_{1}]$ is the 
Schr\"{o}dinger representation of a Markov process. This representation has 
been introduced by Schr\"{o}dinger (\cite{schroe}) and later investigated 
by Kolmogorov and several other mathematicians (for a review in this field 
see (\cite{naga2})). In the Kolmogorov representation, the Markov process 
and its time reversed version define two positive semi-group that are in 
duality relative to the measure $\mu(t,x) dt dx$. The transition 
probability densities $p(s,y;t,x)$ and $\hat p(s,y;t,x)$ verify diffusion 
equations: 
\begin{equation} 
- \frac{\partial p(s,y;t,x)}{ \partial t} + (\hbar /2m) \Delta _{x} 
p(s,y;t,x)+ a(t, x) \nabla _{x} p(s,y;t,x) =0 \label{partial} 
\end{equation} 
and 
\begin{equation} \partial \hat 
p(s,y;t,x) \partial t + (\hbar /2m) \Delta _{x} \hat p(s,y;t,x)+ \hat a(t, 
x) \nabla _{x} \hat p(s,y;t,x) =0 \label{partial} 
\end{equation} 
while the density distribution of the Markov process obeys to the law
\begin{equation}
\frac{\partial \mu(t,x)}{\partial t} + \nabla (\frac{\hat a -  a}{2} 
\mu(t,x)) = 0
\label{evolmu}
\end{equation} 
The drift functions $a$ and $\hat a$ are determined by the 
duality condition, this leads to the celebrated Kolmogorov result 
$a(t,x) = (\hbar /m) \nabla_{x} \ln [\phi (t,x)]$ and $ \hat a(t,x) = 
(\hbar / m) \nabla_{x} \ln [\hat \phi (t,x)]$.\\
Thus the process defined by the entry-exit conditions and the equations
(\ref{dif}) and (\ref{hatdif}) is the Schr\"{o}dinger representation of a 
well defined Markov process. This representation is useful to show the 
existence of a superposition principle for Markov processes. \\ 
Instead of the two real-valued functions $\phi (t, x)$ and $ \hat{\phi} 
(t,x)$, we can introduce two other real-valued functions \begin{equation} 
R(t, x) = \frac{1}{2} \ln{\phi (t, x) \hat{\phi}(t, x)} \quad and \quad
S(t, x) = \frac{1}{2} \ln{\frac{\hat \phi (t, x)}{\phi(t, x)}}
\label{RS}
\end{equation}
that we can combined into one complex-valued function $\Psi (t,x) = exp
[R(t,x) + iS(t,x) ]$. It has been shown (\cite{naga1}) that $\Psi(t,x)$ 
verifies a Schr\"{o}dinger equation
\begin{equation}
i\hbar \frac{\partial \Psi(t,x)}{\partial t} = - \frac{\hbar^{2}}{2m} 
\Delta _{x} \Psi(t,x)  + V(t,x) \Psi(t,x) \label{schro}\end{equation}
in which $V(x, t)$ is related to $u(t, x)$ according to \begin{equation}
V(t, x) - u(t, x) + 2\hbar [\frac{\partial S(t, x)}{\partial t}  + 
D(\nabla S)^{2}(t, x)] = 0 \label{pot}
\end{equation}
Thus, when we go from the representation involving two real-valued 
functions to another one based on one complex-valued function $u(t,x)$ has 
to be changed into $V(t,x)$ as shown by (\ref{pot}). In addition 
to the previous results the equation for the complex conjugate $\bar 
\Psi(t,x)$ of $\Psi(t,x)$ has been established and we have the basic  
result 
\begin{equation}
\mu(t,x)= \phi(t,x) \hat \phi(t, x) = \Psi(t, x) \bar \Psi(t, x)
\label{fifi}
\end{equation}
that gives the physical interpretation of the product $\Psi(t, x) 
\bar \Psi(t, x)$.
Finally, the superposition principle of Markov processes shows that 
$\Psi(t, x)$ verifies the usual superposition principle associated with a 
wave function. Since the Schr\"{o}dinger equation is time reversible in the 
Wigner sense, in this Section we have shown how to implement the 
description presented in previous Sections in order to be 
able to describe reversible processes.\\ 
The derivation of the Schr\"{o}dinger equation that we have developped here 
is totally different from a recent approach in which the complex nature of 
the wave function is connected with the assumption that the space-time, by 
itself, has a fractal nature (\cite{nottale}). Another route to derive the 
Schr\"{o}dinger equation has been developped in a series of very interesting 
papers presented by G.N. Ord (see for instance (\cite{ord1}) and 
(\cite{ord2})). In these papers the role of time-irreversibility is carefully
analyzed and the reversibility of the Schr\"{o}dinger equation is 
obtained by considering more informations on the paths but selecting a 
special projection of the processes that appears as time-reversible. Here 
the route that we have retained is the one that allows to treat on a similar 
footing statistical mechanics and quantum physics.

\section{Concluding remarks}
The existence of a discrete space-time is one of the concepts used to 
describe the physics at the Planck's scale. In this paper 
we show that the existence of a primarily discrete space-time can be also a 
powerful concept to describe the physics in a pre-relativistic world. Our 
work is based on two main assumptions: i) the discrete space-time 
structure is determined by relations that mimic the Heisenberg uncertainty 
relations and ii) the motion in this space-time model is as simple as 
possible. For the properties that we have in mind the continuous limit is 
sufficient. All quantities defined in this approach have a dynamical 
character, they depends on the dynamics on the paths and on the time 
interval, $\tau$, on which we observe the paths. We have shown that it 
exists a characteristic time interval, $\tau =\beta \hbar$, for which the 
mean value of the energy calculated over the closed paths corresponds 
exactly to the energy needed in the system preparation; for this 
like-equilibrium condition our results are identical to the standard ones 
expressed in terms of path integral. We have seen that $\tau$ represents 
the time that we have to wait in order that the quantum fluctuations do not 
exceed the thermal ones. Of course, it is possible to follow the system 
evolution for time intervals smaller than $\tau$ but in this case we have 
not the usual thermodynamics.\\ Our approach relates 
thermodynamics to the existence of motions that are not time-reversible at 
the microscopic level. From this change of paradigm it becomes easy to 
derive a $H-theorem$. To the motion in space-time we associate formal 
collisions, they may appear everywhere and at any time as a consequence of 
the fractal character of the motion. Moreover, in contrast with the 
Boltzmann dynamics, a formal collision does not destroy the correlations 
in the velocities.\\ At least in the simple example investigated here we 
may say that we have demonstrated the Feynman's conjecture $i.e.$ the 
possibility to derive the statistical thermodynamics directly from 
the inspections of the paths and without using all the apparatus of quantum 
mechanics. In this respect there is a parallel with the black hole 
thermodynamics for which it has been shown that the main results can be 
derived from general features of the theory rather than the detailed form 
of the Einstein's equation (\cite{wald}). In addition, it is interesting to 
underline that our $S_{path}$ is basically related to the space-time 
properties, it does not result from the counting of microstates as in 
standard thermodynamics. However, we can perform an exact mapping between 
$S_{path}$ and the thermal entropy, a similar mapping exists between the 
standard heat-bath formalism and the thermal behavior induced by vacuum 
polarization in presence of causal horizons (\cite{schroer}). Of course, 
here we have no quantum vacuum and no horizon and therefore $S_{path}$ has 
not the same origin that the geometrical entropy introduced in the vicinity 
of a black hole, nevertheless from our approach we may emphasize some 
similarities between two different fields of investigations.\\ The results 
obtained in statistical thermodynamics are based on a the properties of a 
real-valued function $\phi(t, x)$ that verifies a diffusion like equation 
(\ref{dif}) but this equation is not a simple imaginary-time version of the 
Schr\"{o}dinger equation, no analytic continuation is invoked in our 
approach. To get the Schr\"{o}dinger equation we force the system to have a 
time-reversible behavior for any time, from this we leave statistical 
thermodynamics and enter in quantum mechanics. 

%%%%%%%%%%%%%%%%%%%%%%%%%%%%%%%%%%%%%%%%%%%%%%%%%%

\end{document}